\documentclass{article}

\usepackage{PRIMEarxiv}

\usepackage[utf8]{inputenc} % allow utf-8 input
\usepackage[T1]{fontenc}    % use 8-bit T1 fonts

\usepackage{url}            % simple URL typesetting
\usepackage{booktabs}       % professional-quality tables
\usepackage{amsfonts}       % blackboard math symbols
\usepackage{nicefrac}       % compact symbols for 1/2, etc.
\usepackage{microtype}      % microtypography
\usepackage{lipsum}
\usepackage{fancyhdr}       % header
\usepackage{graphicx}       % graphics
\usepackage{caption}
\usepackage{siunitx}
\usepackage{makecell}
\usepackage{tabularx}
\usepackage{amsmath}
\usepackage[version=3]{mhchem}
\usepackage{tabularray}
\usepackage{ragged2e}
\usepackage{subfig}
\usepackage{longtable}
\usepackage{float}
\usepackage{pbox}
\usepackage{parskip}
\usepackage{tcolorbox}
\usepackage{hyperref}       % hyperlinks
\graphicspath{{media/}}     % organize your images and other figures under media/ folder
\usepackage{listings}

\title{Fine tuning LLMs for Enterprise: Practical Guidelines and Recommendations}

\author{
  Mathav Raj J\\
  HCLTech\\
  Bengaluru\\
  \texttt{mathavraj.j@hcl.com} \\
   \And
  Kushala VM \\
  HCLTech\\
  Bengaluru\\
  \texttt{kushala.vm@hcl.com} \\
   \And
  Harikrishna Warrier \\
  HCLTech\\
  Bengaluru\\
  \texttt{harikrishna.w@hcl.com} \\
   \And
  Yogesh Gupta \\
  HCLTech\\
  Bengaluru\\
  \texttt{yogeshg@hcl.com} \\
}

\begin{document}
\maketitle

\begin{abstract}
  There is a compelling necessity from enterprises for fine tuning LLMs (Large Language Models) to get them trained on proprietary domain knowledge. The challenge is to imbibe the LLMs with domain specific knowledge using the most optimial resource and cost and in the best possible time. Many enterprises rely on RAG (Retrieval Augmented Generation) which does not need LLMs to be fine-tuned but they are limited by the quality of vector databases and their retrieval capabilities rather than the intrinsic capabilities of the LLMs themselves. In our current work we focus on fine tuning LLaMA, an open source LLM using proprietary documents and code from an enterprise repository and use the fine tuned models to evaluate the quality of responses. As part of this work, we aim to guide beginners on how to start with fine tuning an LLM for documentation and code by making educated guesses on size of GPU required and options that are available for formatting the data. We also propose pre processing recipes for both documentation and code to prepare dataset in different formats. The proposed methods of data preparation for document datasets are forming paragraph chunks, forming question and answer pairs and forming keyword and paragraph chunk pairs. For code dataset we propose forming summary and function pairs. Further, we qualitatively evaluate the results of the models for domain specific queries. Finally, we also propose practical guidelines and recommendations for fine tuning LLMs.
\end{abstract}

% keywords can be removed
\keywords{Fine tuning guidelines \and Code dataset \and Document dataset}

\section{Introduction}
The advent of LLMs has revolutionised natural language processing. Applications are varying from language translation \cite{xu2024paradigm}, content creation \cite{wang2024weaver} and to emotional support chatbots \cite{zheng2023building}. LLMs like LLaMA have been trained on trillions of tokens\cite{touvron2023llama} from various resources. To adapt a general purpose LLM for one of these specific tasks, it has to be trained on task oriented dataset. This additional training allows the model to fine tune its parameters to the task or domain we are interested in. Models like FinGPT for finance domain \cite{liu2023fingpt}, PMC-LLaMA for medical domain \cite{wu2023pmcllama} are fine tuned on particular domain datasets to achieve improved accuracy on domain related questions. Domain specific LLMs can be helpful in scenarios such as support ticket resolution, querying document base or code repository to adapt into new system etc. Though there is an option to use OpenAI models to solve most of the use-cases, there is a high demand for domain specific LLMs due to data privacy and pricing concerns. The stake holder's dataset can stay on premise as the LLMs are also present on premise. Fine-tuned LLMs provide quality and custom feel to the stake holder and also has low latency in displaying the results. 

This paper aims to enable a beginner in preparing the data for fine tuning, estimating the compute capability and memory needed, choosing the right dataset format and the optimal configurations for fine tuning.
The paper is arranged as follows: 
\begin{itemize}
    \item Research Background: A short survey on related research work on fine tuning vs RAG, fine tuning guidelines, efficient techniques for fine tuning and preparation of datasets for fine tuning
    \item Fine tuning configurations: Before starting the fine tuning process it is necessary to understand what configurations can be adjusted to run on available resources.
    \item Proposed Dataset formats for Text and Code: The overall workflow of the proprietary data fine tuning is detailed in this section. The different proposed formats of text and code data has been explained in detail
    \item Experiments: A proprietary document and code repository is used to showcase the fine tuning work flow. Empirical studies have been conducted to understand the effects of quantization on time and memory, to understand the selection of appropriate rank in LORA (Low Rank Adapater) fine tuning, understand the memory requirements for full model fine tuning and to understand the effects of fine tuned model in a Retrieval Augmented Generation (RAG) pipeline
    \item Guidelines: In the final section, practical guidelines for fine tuning has been given. Some tips to choosing right parameters for fine tuning efficient techniques like LORA is also listed.
\end{itemize}

\section{Research Background}

The authors in \cite{app14052074} give an overall view of the recent trends in LLMs. LLMs can be trained in different phases. The first phase being the pretraining with defined objectives such as causal language modelling, masked language modelling, span denoising objective etc. Then comes the transfer learning phase which is further classified as feature based transfer and finetuning approach. Transfer learning is required when the dataset is inadequate for a complete training from scratch. Therefore pretrained weights are used as starting point. In feature based transfer learning features obtained from the pretrained model for the given domain dataset are used by another smaller model to train. Meanwhile finetuning on a dataset is to nudge the pretrained weights on a particular task oriented dataset. Depending on the how many layers are fine tuned and how prompts are handled during finetuning, it is further classified as adapter tuning, gradual unfreezing, prompt tuning etc. 

An alternate approach to fine tuning is to chunk the documents, convert to embeddings and store it in a vector database for retrieval using similarity search with the query and use the pretrained LLM to come up with a consolidated answer from the retrieved documents. \cite{Jeong_2023}.  This is the production ready approach as it is fast and gives more or less exact results. However the RAG approach can be crippled by a not so good retrieval mechanism. Though there are simple alleviations like the claim in \cite{cuconasu2024power} that information retrieval in RAG has improved by purposeful addition of noisy documents, the quality of answers are limited by the similarity search which has nothing to do with the LLM capability itself. 

A study by \cite{balaguer2024rag} shows the comparison of between finetuning and RAG. The experiments in the paper reveal that finetuning on a domain data extracted from agriculture journals have given more succinct and accurate responses than a RAG pipeline. That being said, the authors also have ackowledged the high initial cost required to fine tune a model.
 
Low Rank Adaptation (LORA) finetuning of LLMs \cite{hu2021lora} has opened up a whole new possibility of finetuning limited number of essential parameters usually of the order of few thousands to a millions instead of the entire parameters which is in the order of billions. The work on quantizing LLMs has opened up avenues for resource deficient systems to train at low memory cost \cite{dettmers2022llmint8}. Papers on quantized LLMs in combination with parameter efficient techniques like LORA have further enabled obtaining satisfactory results with low resources\cite{NEURIPS2023_1feb8787}.

Code generation with LLM is the most attractive task engineers are looking at. Even though LLM are already trained with lots of data which makes them to generate code depending on the input, the challenge is generating code about a specific enterprise domain code. LLM fine-tuning for a specific task makes the model utilize its capacity to fullest by making it adapt to a domain by making the model familiar to jargons, domain terminology by understanding the context of the code, class, functions, exceptions, libraries etc., Fine-tuning also helps in adapting the model to address the task specific problems. Fine-tuning large language models for code-related tasks presents a myriad set of challenges that must be carefully addressed to ensure optimal performance and reliability. The challenges encompass aspects such as data quality and quantity, domain-specific understanding, tokenization and vocabulary, contextual understanding, code generation quality, code understanding vs. generation, model size and computational resources, overfitting and generalization, evaluation metrics, and ethical and security concerns.

\section{Fine Tuning LLMs on Available Resources}
\subsection{Quantization}
By default, most open source LLM weights are released in full 32 bit floating point precision. Even for fine tuning a model of relatively smaller size say 7 billion parameters, nearly 28 GB space is required. With higher precision weights, compute units have to spend higher energy in memory movement operations during fine tuning \cite{6757323}. Quantization is the process of constraining an input from continuous set of values to a discrete set. Quantizing the model weights to a lower precision and fine tuning greatly reduces the size without hampering the quality \cite{dettmers2022llmint8}. 

Data is stored in different numeric types namely: FP32, FP16, int8, FP8, BF16. Integers can be represented in unsigned or signed format or 2's complement form. To represent a fractional decimal we could go for a fixed point representation. To further extend the range, systems use the IEEE 754 floating point representation which has a much larger range since the numbers are expressed in exponents of 2. The 32 bits of floating point representation has three parts namely 1 sign bit, 8 bits for exponent (both positive and negative) and 23 bits for mantissa or significant figures. The width of the exponent bits determines the range of numbers and the width of the mantissa bits determines the precision of numbers. Based on the widths there are different forms. Different forms may be needed based on the availability of memory resources. FP16 reduces both the range and precision by using 5 bits for exponent and 10 bits for mantissa. Brain float 16\cite{kalamkar2019study} or BF16 maintains the same range as FP32 but reduction of precision to 7 bits. Floating point 16 or FP16 enables training of larger models or training with larger mini-batches.

With neural networks, quantization can be done with the weights during storage and activation during the computation. Quantization  (QAT) and post training quantization (PTQ). Integer quantization favours memory reduction and thereby energy and cost reduction during inference \cite{6757323}. In a QAT scheme proposed by Jacob et al. \cite{jacob2017quantization}, the quantization errors are included in the computation graph during fine tuning so that the model's inference performance can be as if it were never quantized. This allows for deployment of models in edge hardware that support only integer arithmetic.

Tim Dettmers et al. propose a new type of integer quantization scheme called LLM int8 \cite{dettmers2022llmint8}. This has been implemented in the python library 'bitsandbytes'. To explain in a more detailed manner, quantization is a two step process that involves i) Finding the normalization constant and scale the vector into target range ii) Rounding off to the nearest value in target range. During matrix multiplication of tensors, quantization of weights with outliers will lead to huge quantization loss. To mitigate this, bitsandbytes employs a combination of vector wise quantization and mixed precision decomposition to achieve a performance similar to that without quantization. Though LLM int8 quantization does not degrade performance, the inference time gets increased due to the overhead of quantization \cite{infissue}. However the memory gets reduced drastically by 71\%, which is major cost saving when choosing cloud premise GPUs.

\subsection{Gradient Accumulation}
Other than quantization, techniques like gradient accumulation help in reducing the memory requirement during fine tuning. The process of accumulating gradients in the context of backpropagation involves a strategic approach to parameter updates within a neural network during the training phase. Unlike the conventional method where parameters are updated after processing each mini-batch, the accumulation of gradients entails deferring the parameter updates until all the instances in a mini-batch have been processed. 

In the standard back propagation algorithm, the gradients computed for each instance in a mini-batch are typically used to immediately update the model parameters. However, in the case of accumulated gradients, these individual gradients are not immediately applied to the parameters. Instead, they are summed or averaged over the entire mini-batch. As each instance in the mini-batch undergoes forward and backward passes, the gradients with respect to the model parameters are computed but not immediately applied. These gradients are stored, and the accumulation occurs over the entire mini-batch. Only when all instances in the mini-batch have been processed, the accumulated gradients are employed to update the model parameters. This aggregated update is akin to the effect of utilizing a higher batch size for training the neural network.

\subsection{PEFT (Parameter Efficient Fine Tuning)}

Large language models get more efficient with transfer learning through fine tuning. But on the other hand, fine tuning becomes challenging with respect to the infrastructure needed, time required and overall memory needs. To overcome these challenges, parameter efficient fine tuning comes into picture. Parameter-efficient Fine-tuning (PEFT) is a technique used in Natural Language Processing to improve the performance of pre-trained language models on specific downstream tasks. It involves reusing the pre-trained model’s parameters and fine-tuning them on a smaller dataset, which saves computational resources and time compared to training the entire model from scratch. PEFT achieves this efficiency by freezing some of the layers of the pre-trained model and only fine-tuning the last few layers that are specific to the downstream task. There are many methods in PEFT training: Adapter, LoRA, QLoRA, Prefix tuning, Prompt tuning, P-tuning, and IA3. In this paper, we will be delving on LoRA and QLoRA methodology of training. 

LoRA, \cite{hu2021lora} a method for fine-tuning large language models, operates by integrating small trainable submodules alongside feed-forward layers in the pre-trained transformer architecture. These LoRA modules employ rank decomposition to significantly reduce the number of trainable parameters while maintaining or even enhancing model performance across various tasks. Specifically, LoRA inserts two feed-forward layers adjacent to each feed-forward layer in the transformer model, where the first layer projects the input into a lower-dimensional space and the second layer restores it to the original dimensionality. This incremental change, represented as delta h, is added to the original hidden representation, resulting in an updated representation h'. Through this approach, task-specific parameters are minimized, facilitating efficient task-switching and reducing hardware requirements without introducing additional inference latency.  

QLoRA, or quantized LORA, is an optimized version of LoRA, where the precision of the weight parameters are reduced to 4 bit precision. Since QLoRA shrinks the model size due to the reduced precision, it is helpful in scenarios where there is limited memory to fine tune.

\section{Fine Tuning Workflow}

The workflow for fine tuning an LLM, be it for text or code, can be represented as in Figure 1 below. 

\begin{figure}[ht]
  \centering
  \includegraphics[width=\linewidth]{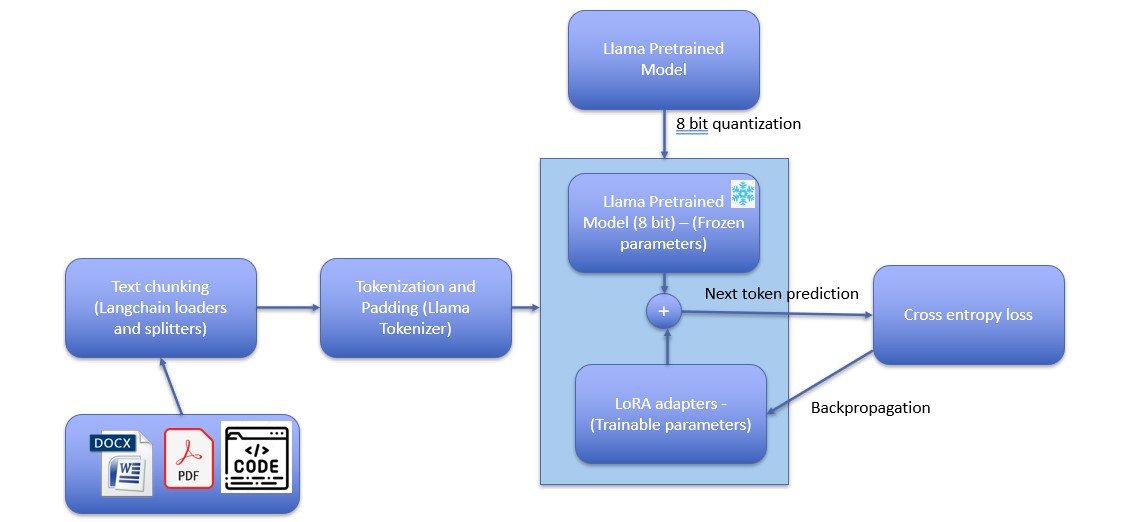}
  \caption{Fine tuning Workflow (with LLaMA model as an example)}
\end{figure}

Unstructured text or code is fed to the first stage of fine tuning, the pre-processing recipe. Based on the recipe, it chunks the data into logical parts and then it is tokenized (and padded where needed) so that it accomodates the supported sequence lenght of the model. Further, the LoRA / QLoRA configurations are applied and fine tuning is done till the loss is minimized. This is the standard process of fine tuning. Thus, we can see that the key step in fine tuning is the data pre-processing, and the quality of fine tuning is primarily dependent on that. So, we will describe that in more detail below.

\subsection{Text Data Pre-Processing}

This step plays a vital role in fine tuning process. From scratch, a Large language model is trained to do next token prediction. The datasets are massive text corpus like Common crawl, The Pile and code repositories from Github. Further the model can be fine tuned for specific tasks using specialised datasets. One such series of datasets are called Instruction datasets like Dolly, Orca, Alpaca and Vicuna. All these datasets enable instruction fine tuning. Instruction fine tuning bridges the gap between next token prediction and user's requirement of following instruction prompts\cite{zhang2023instruction}. An instruction dataset consists of instruction prompt, response paired with or without context for additional information. The challenge is how to create an instruction dataset from a general document. 

As part of our experiments, we developed fine tuning datasets pre-processing recipes of four different formats namely a) raw data b) keywords as instruction c) headings as instruction and d) queries as instruction. 

For a decoder model like LLaMA 2, the raw format is essentially continuing the initial training objective of unsupervised next token prediction with raw text. In the raw data method, we are passing the document as raw chunks with no template. 

In the keywords method, we are passing the chunks in a template with prompt being keywords extracted from the chunk. For this, we used Rapid Automatic Keyword Extraction (RAKE) algorithm (the corresponding python library being rake-nltk\cite{rake}). 

Another method is to use the document headings with different headings levels as prompt. As an example, in cases of Microsoft Word docx documents, the document parser from llm search\cite{llmsearch} is used. 

The last methods that we used is a query based approach. This is particularly usefuly when a user is interested in querying for information about the document from the LLM, hence it is ideal if the prompt is also having queries. The same model is used to generate possible queries that can be asked about a chunk and then the dataset is prepared. A single chunk can have multiple queries to promote diversity in the dataset.

When data set is not in a paired format, the challenge is how to structure the data and let the model understand the given text. As a baseline, we split the text into chunks of fixed length and trained the model with those chunks. The objective of fine tuning the model is next token prediction. As a first step the model encodes the splitted chunks into embeddings.  LLaMA tokenizer is a Byte Pair Encoding tokenizer model based on sentencepiece. Once the words are tokenized, embedded and padded to appropriate fixed sequence length, the data is ready to be fed to model for fine tuning.

In Fine tuning a paired dataset format, the input ids will be the input+output+pad tokens and labels will be the ignore tokens+output+pad tokens. Here we have essentially masked the length of input tokens and it will not be considered while calculating the loss. The input ids and labels for the transformer are the same except that in case of input ids the padding token is <unk> and for labels it is -100. 

The drawback in this approach is that the structure in the document such as headings and topical information might not be present within a chunk. The overlap option in Langchain splitters helps with this issue to an extent. However some topical information are too long to be captured by just overlapping.

\subsection{Code Data Pre-processing}

LLM models are trained such that higher the context, detailing and prompting higher the results. When LLM are trained for code generation task it is very important to make the model understand the domain with quality along with quantity information. Researchers have been exploring the best possible way to reduce the amount of effort required during the preparation of the data. 

In our experiment we have considered three different ways of preparing the training data. They are a) Summary method b) Metadata method and c) Tokenization method

The first method involves splitting the code at a class level or functional level code. The functional level code is considered as a source for preparing the data. The entire code repository is split into function level code. The functional level code is fed into the instruct model to generate the summaries by prompting the model. This type of data becomes our generated dataset which has function level code associated with their summaries.

The second method involves extracting information from the coding practices embedded in the code. It is said that synthetic, structured, high quality, text book like data makes LLM learn faster and produce good results\cite{gunasekar2023textbooks}. In this approach, the comments and docstrings in the code are extracted along with detailed comments and is used along with the raw information gathered from the code as pre-processing data. 

The third method involves tokenizing the whole code base irrespective of the file type into the supported sequence length. This method doesn’t involve gathering any other data. The LLM model with this tokenized data is trained for the purpose of next token prediction usecase.

\subsection{Compute Estimation} \label{cest}

After deciding on the input, the expected output and objective at hand, the next step is to decide on the hardware. To get the memory that will be occupied by a model in FP32 precision a rough estimate is to multiply the model parameter size by 4, since a single parameter occupies 4 bytes. For FP16, the multiplication factor is 2, and for 8 bit quantized model it is 1 and hence for 4 bit it is 0.5. This estimate alone will not help us in finding the right hardware for fine tuning as a higher percentage of memory is required to store the gradients and activations during the fine tuning process. This additional storage also has to be accounted for.

\section{Experiments}

LLaMA 2, the open source model used for experimentation in this paper has a pre training data cut off time on September 2022. Additional data till July 2023 has also been added as fresh data during fine tuning before its release \cite{touvron2023llama}. Hence to test the approaches of data preparation recipes, proprietary documents and code repositories were used from our inhouse machine learning platform.

A100 80 GB Nvidia GPU from Google Cloud Platform is used as the hardware accelerator for the fine tuning process. 

\subsection{Text}
To fine tune with text data, documentation resources from our machine learning platform were used. The user guide of this platform which has 66 pages and 5 MB of data was used as the raw source of data. This was converted to 60 KB csv file with 33 rows. Since the document is not available to public and was also created after the fine tuning cut off time of the original LLaMA 2 model, it makes a good dataset to study. The user guide is in PDF format and the content is well structured with index, headings and step by step instructions. 

\begin{figure}[ht]
  \centering
  \includegraphics[width=0.45\columnwidth]{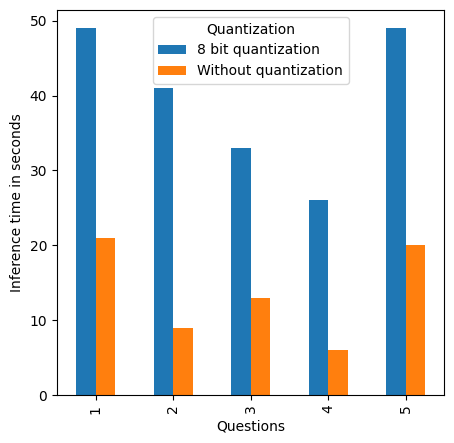}
  \caption{Inference time of fine tuned Llama 2 7B Chat model}
  \label{fig:quantlatency}
\end{figure}

Before the start of fine tuning process, the influence of quantization on inference is shown in table \ref{quanttable}. Quantizing a model saves on GPU memory and allows fine tuning with higher batch sizes there by reducing time and money spent on the training job. However as shown in figure \ref{fig:quantlatency}, quantized model on average takes a higher time for inference compared to a non quantized model. 

\pagebreak

\begin{table*}[ht]
\centering
\smallskip
\captionof{table}{Influence of quantization on inference}
\label{quanttable}
\begin{tabularx}{\textwidth}{|l|l|X|}\hline Quantization  & GPU memory & What is AION?\\
\hline Without quantization & 28 GB &\RaggedRight{AION (Artificial Intelligence ON) is a cloud-based platform that enables developers to build, train and deploy machine learning models. It provides an end-to-end solution for data scientists and engineers to create, test, refine, and deploy ML models in production environments.}\\
\hline 8 bit quantization & 8 GB &
\RaggedRight{AION (Artificial Intelligence ON) is a cloud-based platform that enables developers to build, train and deploy machine learning models. It provides an end-to-end solution for data scientists and engineers to create, test, refine, and deploy predictive modeling solutions in the form of APIs or containerized microservices.}\\
\hline
\end{tabularx}
\end{table*}

In table \ref{tab:llamatable}, the maximum possible configurations of PEFT methods on a A100 80 GB GPU machine are listed. For example, 70 B parameter flavour of LLaMA 2 model can be fine tuned only with QLoRA in 80 GB machine. LORA is not possible due to memory constraint.

\begin{table*}[ht]
\centering
\smallskip
\begin{talltblr}[
caption = {Maximum possible PEFT configurations of Llama 2 models on A100 80 GB},
  label = {tab:llamatable},
                ]{hlines, vlines,
                  colspec = { l *{9}{c}}
                  }
    \thead{Model \\ \& size}  & \thead{Dataset size}& \thead{Epochs}  & \thead{PEFT \\ method} & \thead{CPU \\memory} & \thead{GPU \\memory} & \thead{Estimated \\time}\\
        Llama 2 Chat 7B  & 60 KB & 3 & LORA & 6 GB  & 18 GB
        & 15 mins\\ 
        Llama 2 Chat 13B  & 60 KB & 3 & LORA &  6 GB & 26 GB
        & 25 mins\\ 
        Llama 2 Chat 70B & 60 KB & 3 & QLORA & 7 GB   & 65 GB
        & 40 mins\\  
\end{talltblr}
    \end{table*}

Now with the understanding of PEFT configurations possible for different size models, next there is a need to tune the PEFT hyper parameters. In \cite{hu2021lora}, alpha is suggested to be fixed and rank is fine tuned. Typically lower ranks are preferred with a number of target modules. However owing to a smaller dataset size, target modules were kept as q\_proj and v\_proj only. A high alpha is used so that model could learn more from new gradient updates pertaining to the new information from documentation. The quality of the responses are captured with a manual assessment. As evident from table \ref{lorar}, the appropriate rank and alpha that gives decent results vary for the different parameter sized models.

\begin{table*}[ht]
\centering
\smallskip
\captionof{table}{LoRA rank tuning}
\label{lorar}
\begin{tabular}{l p{1cm} *{8}{p{1.3cm}}}
\toprule
\multicolumn{1}{c}{Model and size} &
  \multicolumn{2}{c}{LoRA parameters}  &
  \multicolumn{2}{c}{Prompt Manual assessments}\\
\cmidrule(lr){2-3} \cmidrule(lr){4-5}  
  & {Rank} & {Alpha}  & {Qn1} & {Qn2}\\
\midrule
7B Chat  & 1 & 512  & Good & Good \\
     & 2 & 512  & Average & Poor \\
      & 4 & 512  & Poor & Poor \\
    & 8 & 512  & Poor & Poor \\
    & 16 & 512  & Poor & Poor \\
\midrule
13B Chat  & 1 & 512  & Poor & Good \\
     & 2 & 512  & Poor & Average \\
      & 4 & 512  & Poor & Good \\
    & 8 & 512  & Good & Good \\
    & 16 & 512  & Average & Average \\
\bottomrule
\end{tabular}
\end{table*}

With data extracted from the pdf, the final dataset is prepared in four ways and it is clear that each variation of the data preparation recipe produces different small changes in the response. 'Raw' and 'Keyword' recipe give short and succinct responses. 'Heading' and 'Summary' methods give a somewhat detailed response.

\begin{table*}
\centering
\smallskip
\captionof{table}{Influence of data preparation recipes on inference}
\label{datar1}
\begin{tabularx}{\textwidth}{|l|X|}

\hline Dataset type  & What is AION?\\
\hline Default & \RaggedRight{ AION (Artificial Intelligence ON) is a cloud-based platform that enables developers to build, train and deploy machine learning models. It provides an end-to-end solution for data scientists and engineers to create, test, refine, and deploy predictive modeling solutions in the form of APIs or containerized microservices. }\\
\hline ML &
\RaggedRight{AION (AI-ON) is an open source platform for building, deploying and managing machine learning models at scale. It provides features such as data version control, reproducibility, collaboration support, model monitoring, and deployment to cloud or on premise infrastructure.}\\
\hline Meta Data &
\RaggedRight{AION (Artificial Intelligence ON) is an open-source, cloud-native platform for building and deploying modern machine learning models at scale. It provides a unified environment to develop, train, test, and deploy ML models across different data sources and platforms. The main goal of AION is to make it easy for developers to build, manage, and monitor their ML workflows by automating most of the underlying tasks such as data ingestion, feature engineering, model training,} \\
\hline Gen AI &
\RaggedRight{AION (Artificial Intelligence ON): It refers to the integration of machine learning, data analytics, robotics, and artificial intelligence to automate business processes, improve productivity, and enable more informed decisions.AION stands for Artificial Intelligence ON which means all the features such as ML, Deep Learning, etc are enabled by default without any need to disable them. This makes it easy to use and deploy models in production environments seamlessly.}  \\
 \hline
\end{tabularx}
\end{table*}

\begin{figure}%
    \centering
    \subfloat[\centering 7B Chat model full fine tuning]{{\includegraphics[width=5cm]{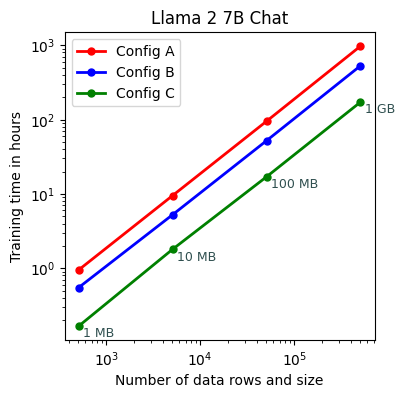} }}%
    \qquad
    \subfloat[\centering 13B Chat model full fine tuning]{{\includegraphics[width=5cm]{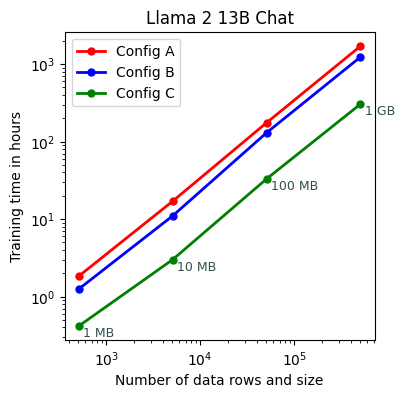} }}%
    \caption{Fine tuning time of Llama 2 Chat models on different fine tuning configurations}%
    \label{fig:fftplots}%
\end{figure}

PEFT methods are recommended for smaller dataset. However if there is a huge collection of dataset with new information, full fine tuning of all the parameters of a model is possible. The A100 80 GB GPU is capable of supporting full model fine tuning of 7B and 13B models. In table \ref{tab:fulltable} the loading parameters and batch size details are given in detail. Higher batch sizes gave cuda out of memory error. The configurations in each row of each of the model size are named as config A, B and C respectively. In \ref{fig:fftplots}, a graph is plotted between number of data rows and the training time. With increase in data rows, the time increases linearly. A different configuration with higher batch size and gradient accumulation steps will decrease the fine tuning time slightly. A huge time will be saved if full fine tuning is done for half the total capacity of maximum sequence length of LLaMA 2 models.
\begin{table*}[ht]
\centering
\smallskip
% \captionof{table}{Full fine tuning of LLaMA 2 Chat on A100 80 GB}
% \label{fulltable}
\begin{talltblr}[
caption = {Full fine tuning of Llama2 Chat on A100 80 GB},
  label = {tab:fulltable},
                ]{hlines, vlines,
                  colspec = { l *{9}{c}}
                  }
    \thead{Llama2 Chat \\Model \\ size} & \thead{Data size}  & \thead{Seq \\len}
        & \thead{Model \\precision} & \thead{Batch \\size}  & \thead{Gradient \\accumulation \\steps} & \thead{Resulting \\steps}  & \thead{CPU \\memory} & \thead{GPU \\memory} & \thead{Estimated \\time\\(mins)} \\
        7B  & 60 KB & 4096   & FP32
        & 1   & 2  & 254 
        & 54 GB  &  80 GB    & 57                     \\ 
       7B   & 60 KB & 4096   & FP16
        & 2   & 4  & 63 
        & 54 GB  &  80 GB    & 33                     \\ 
        7B    & 60 KB & 2048   & FP16
        & 4   & 8  & 16 
        & 54 GB  &  80 GB    & 10                     \\ 
        13B    & 60 KB & 4096   & FP32
        & 1   & 2  & 254 
        & 102 GB  &  80 GB    & 110                     \\ 
        13B   & 60 KB & 4096   & FP16
        & 2   & 4  & 127 
        & 102 GB  &  80 GB    & 75                    \\ 
        13B   & 60 KB & 2048   & FP16
        & 4   & 8  & 16 
        & 102 GB  &  80 GB    & 25                     \\ 
\end{talltblr}
    \end{table*}

RAG- (Retrieval Augmented Generation) is a popular way of using embeddings and a vector similarity search to find only relevant context from a pool of documentation with new information or any content longer than maximum capacity of the LLM. The LLM interprets the context and gives a polished response. This entire process can be summed as embedding conversion followed by similiarity search usually through a vector DB which in turn is followed by LLM handling the context. It is evident from this process that the contribution of LLM will be good only if the contexts obtained are good. A comparison is made in table \ref{tab:ragvsft} between RAG with a pretrained model and RAG with the fine tuned model. It is seen that RAG responses are improved with specific answers and even following the style of the document. Since the document being dealt here is a user guide there are steps mentioned in every process. Only the fine tuned model was able to follow this pattern of giving responses in step by step format. Another observation is regarding the last question which is a multi part question.  The last question was not answered by the base LLaMA 2 model. Our fine tuned model was not only able to follow the style but also give the precise answer to the second part of the question. Hence fine tuning could be one solution to hallucinations in RAG.

\pagebreak

\begin{longtable}{| p{.2\textwidth} | p{.35\textwidth} | p{.37\textwidth} |}

\caption{Retrieval Augmented Generation with LLaMA 2 Chat before and after fine tuning} % needs to go inside longtable environment
\label{tab:ragvsft}
\\
\hline
Questions  & RAG & Finetuned model and RAG  \\
\hline 
What is AION? & \RaggedRight{Based on the provided context, I can confirm that AION is an artificial intelligence (AI) life cycle management platform designed to help organizations apply machine learning to real-world problems. AION provides a comprehensive suite of tools and services for managing the entire AI lifecycle, including data ingestion, exploratory data analysis, data processing, feature selection, model training and hyperparameter tuning, and inference services. The platform offers a less-code/no-code approach, allowing business users to generate ML models without coding. Additionally, AION includes explainability capabilities to provide higher accuracy predictions with supported explanations and model confidence.} & \RaggedRight{Based on the provided context, I can explain what AION is.

AION stands for Artificial Intelligence ON, which is an AI life cycle management platform used for solving real-world problems using machine learning. AION provides a less-code/no-code approach to users, allowing them to train and deploy machine learning models without any coding knowledge. It includes various sub-processes such as AutoML, MLOPs, MLaC, and Explainability. These sub-processes are automated, and AION generates the best deployment model after analyzing the input data and problem statement.} \\ \hline 
\RaggedRight{Write in detail about MLLite feature in AION. List the input model formats that can be converted to ONNX.} & \RaggedRight{Based on the provided context, here is the information regarding the MLLite feature in AION:
MLLite is a feature in AION that allows users to convert their trained models into formats compatible with various deep learning frameworks, including ONNX and TFLite. The main purpose of MLLite is to enable efficient inference execution on different devices without requiring significant modifications to the original model architecture.
Regarding the input model formats that can be converted to ONNX, according to the context, the following formats are supported:Keras, Sklearn, TensorFlow, TFLite, Light GBM, XGBoost, LibSVM,CoreML
In summary, MLLite is a useful feature in AION that enables users to convert their trained models into optimized formats for efficient inference execution on various devices, while also providing support for multiple input model formats for conversion to ONNX.}  & \RaggedRight{Sure! Here’s what I could gather from the given text regarding the MLLite feature in AION:MLLite feature in AION allows users to convert their trained models into different formats such as ONNX, TFLite, and Sagemaker. The main advantage of this feature is that it enables users to consume these optimized models for predictions without having to retrain them.Input Model Formats that can be converted to ONNX include:Keras,Sklearn,TensorFlow,Light GBM,XGBoost,LibSVM, CoreML}\\ \hline
\RaggedRight{Is there Incremental learning in AION? If so what are the steps to create an incremental model?} & \RaggedRight{ Yes, AION supports incremental learning for regression and classification problems. To create an incremental model in AION, follow these steps:
1. Select the "Online Learning" (Beta) or "Distributed Learning" (Beta) checkbox in the Incremental Learning section of the configuration page.
2. Choose the problem type (Regression or Classification) that supports incremental learning.
3. Proceed to the next step by clicking the "Next" button.
In summary, to create an incremental model in AION, you need to enable online learning in the configuration page and choose the appropriate problem type.}  & \RaggedRight{Yes, there is incremental learning in AION. To create an incremental model follow the below-mentioned steps:

Step 2: Click online learning tab.

Step 3: Select the Online Learning (Beta) or Distributed learning (Beta) checkbox for Incremental Learning.

Step 4: Click next icon to proceed to advanced configuration tab.}\\ \hline
\RaggedRight{How can we do text labelling with AION? Give the API URL for model monitoring?} & \RaggedRight{Based on the provided context, the API URL for model monitoring during text labeling with AION is not explicitly mentioned. Therefore, I cannot provide an answer to this question.}  & \RaggedRight{Based on the provided context,

I can answer the question as follows:

To perform text labelling with AION, follow these steps:

Step 1: Go to 

Utilities > Text Data Labelling.

Step 2: Upload the text data in 

CSV format.

Step 3: Select relevant features and click Next.

Step 4: Generate predictions by clicking Predictions tab.

The predicted labels are displayed under Predicted Labels section.

For example, if the input text is "This is a sample text", the predicted label would be "positive".

API URL for Model Monitoring:

To monitor the performance of the model, users can use the API endpoint /monitoring/model-performance. This API returns the summary statistics of the model’s performance along with the confusion matrix.

Example:

curl -X POST 

http://localhost:8080/monitoring/model-performance -H 'Content-Type:

application/json' -d '{"data":["This is a sample text"],"label":["positive"]}'}\\ \hline

\end{longtable}
% \pagebreak

Finally the objective of the fine tuning process is to see if the LLM has learnt the new knowledge and as shown in table \ref{aiontable}, it indeed has learnt.

\begin{center}
\smallskip
\captionof{table}{Model learning new information after fine tuning}
\label{aiontable}
\begin{tabularx}{\textwidth}{|l|l|X|}
\hline Fine tuning  & Model & What is AION?  \\
\hline Before & Llama 2 chat &\RaggedRight{Aion (AION) is a blockchain-based platform that enables the creation and exchange of digital assets, such as NFTs. It was founded in 2018 by a team led by Matthew Roszak, who has extensive experience in the cryptocurrency industry. } \\
\hline After & Llama 2 chat &
\RaggedRight{AION (Artificial Intelligence ON) is a cloud-based platform that enables developers to build, train and deploy machine learning models. It provides an end-to-end solution for data scientists and engineers to create, test, refine, and deploy ML models in production environments. } \\
\hline
\end{tabularx}

\end{center}

\subsection{Code}

% \pagebreak
% \clearpage

For the code dataset, JAVA files from a sustenance engineering solutions platform was used. This being a closed source code repository, the codes are new unlearned information for Llama models making it a good dataset to experiment with. The total size of the code data was 16MB with a total function count of 13807. The dataset was condensed and packed according to token limit of LLaMA model and resulted in a csv file of 10 MB size with 60 rows. 

\begin{table}[ht]
\centering
\smallskip
% \captionof{table}{Code LLaMA PEFT}
\label{codet1}
\begin{talltblr}[
caption = {PEFT methods on Code Llama on A100 80 GB},
  label = {tab:codet1},
                ]{hlines, vlines,
                  colspec = { l *{9}{c}}
                  }
    \thead{Model \\ \& size} & \thead{Data size} & \thead{Epochs} & \thead{PEFT \\ method} & \thead{CPU \\memory} & \thead{GPU \\memory} & \thead{Estimated \\time} \\
        Code Llama 7B & 10 MB & 3 & LORA & 6 GB  & 18 GB
        & 23 mins                       \\ 
        Code Llama 13B  & 10 MB & 3 & LORA &  6 GB & 26 GB
        & 25 mins                       \\ 
        Code Llama 7B & 10 MB & 3 & QLORA & 7 GB   & 15 GB
        & 10 mins                       \\  
\end{talltblr}
\end{table}

Table \ref{tab:codet1} exhibits the experiment on the hyper parameters conducted to investigate the performance of the model. Few results are discussed by prompting the trained model and comparing the results from the code repository. It was seen from the results from the experiments the Code LLaMA models gave an exceptionally good results with LoRA rank 16 and alpha 32. Higher the rank and alpha made the models to hallucinate and randomly generate the code.

  \begin{table*}[ht]
\centering
\smallskip
% \captionof{table}{Code LLaMA full finetuning}
\label{datar2}
\begin{talltblr}[
caption = {Full fine tuning of Code Llama on A100 80 GB},
  label = {tab:?},
                ]{hlines, vlines,
                  colspec = { l *{9}{c}}
                  }
    \thead{Model \\ \& size}  & \thead{Seq \\len}
        & \thead{Model \\precision} & \thead{Batch \\size}  & \thead{Gradient \\accumulation \\steps} & \thead{CPU \\memory} & \thead{GPU \\memory} & \thead{Estimated \\time} \\
        Code Llama 7B    & 4096   & FP16
        & 2   & 2  
        & 54 GB  &  57 GB    & 10 mins                   \\ 
        Code Llama 7B    & 4096   & FP16
        & 2   & 4  
        & 54 GB  &  46 GB    & 38 mins                     \\ 
        Code Llama 7B    & 2048   & FP16
        & 4   & 8   
        & 54 GB  &  59 GB    & 10 mins                     \\ 
        Code Llama 13B    & 4096   & FP32
        & 2   & 2   
        & 102 GB  &  53 GB    & 9 mins                     \\ 
        Code Llama 13B   & 4096   & FP16
        & 2   & 4   
        & 102 GB  &  53 GB    & 18 mins                     \\ 
        Code Llama 13B   & 2048   & FP16
        & 4   & 8   
        & 102 GB  &  62 GB    & 33 mins                     \\ 
\end{talltblr}
    \end{table*}

 \begin{table}[ht]
\centering
\smallskip
\captionof{table}{Code Llama LORA}
\label{datar3}
\begin{tabular}{l p{1cm} *{6}{p{2cm}}}
\toprule
\multicolumn{1}{c}{Model and size} &
  \multicolumn{2}{c}{LoRA parameters}  &
  \multicolumn{2}{c}{Prompt Manual assessments} \\
\cmidrule(lr){2-3}  
  & {Rank} & {Alpha}    \\
\midrule
Code Llama 7B  & 2 & 4  & Poor  \\
     & 4 & 8  & Poor  \\
      & 8 & 16  & Average  \\
    & 16 & 32  & Good  \\
\bottomrule
\end{tabular}
\end{table}

\clearpage
\noindent
\begin{center}
\begin{longtable}{| p{.2\textwidth} | p{.35\textwidth} | p{.37\textwidth} |}

\caption{Database connection code} % needs to go inside longtable environment
\label{tab:dbcode}
\\
\hline
Prompt  & Ground truth & Predicted response  \\
\hline

A connection object is created to connect to a database for a given project name

Connection is established to the database using the URL, username and password

If an exception occurs the error message is printed
The connection object is returned to the calling function & \RaggedRight{\begin{lstlisting}
Connection con = null;
HttpServletRequest httpRequest = null;
try{
System.out.println(""get New Connection projectName:""+projectName) ;
Class. forName(driverName) .newInstance() ;
con = DriverManager.getConnection(url+projectName, dbUser, dbPass);
} catch (SQLException sqle) {
System.out.println(""in ConnectionPool of getNewConnection()11111""+sqle.getStackTrace());
} catch (Exception e) {
System.out.println(""in ConnectionPool of getNewConnection()222"");
e.printStackTrace();
}
return con; 
\end{lstlisting}} & \RaggedRight{\begin{lstlisting}
public Connection getConnection(String projectName){
Connection con = null;
HttpServletRequest httpRequest = null;
try{
    Class.forName(driverName).newInstance();
    con = DriverManager.getConnection(url+projectName, dbUser, dbPass);
} catch (SQLException sqle) {
    System.out.println("In ConnectionPool of getConnection()"+sqle.getMessage());
}} catch (Exception e) {
    System.out.println("In ConnectionPool of getConnection()"+e.getMessage());
}
return con;
\end{lstlisting}}\\ \hline 
\end{longtable}
\end{center}

\clearpage
\noindent
\begin{center}
\begin{longtable}{| p{.2\textwidth} | p{.35\textwidth} | p{.37\textwidth} |}

\caption{Code for getting testcase by ID} % needs to go inside longtable environment
\label{tab:tstcode}
\\
\hline
Prompt  & Ground truth & Predicted response  \\
\hline

Connect to the database.

Prepare a query with a given test case ID to the table testcases table.

Set the parameter to the given value.

Execute the query and retrieve the results.

Get the desired data from the results.

Release resources used by the query and results.

Return the desired data. & \RaggedRight{\begin{lstlisting}
public  List<DefectTestCaseMappingDTO>  getFieldDefects(String testcaseId){
int tcIdPk = DeltaUploadDAO.getInstance().getTestCaseId(testcaseId);
List<DefectTestCaseMappingDTO> fieldDefectsList = new ArrayList<DefectTestCaseMappingDTO>();
Connection connection = ConnectionPool.getInstance().getConnection();
if(null != connection){
Statement  stmt = null; 
ResultSet rs = null;
try {
if(null != connection){
    String sql = ""SELECT * FROM DEFECT_TESTCASE_MAPPING  WHERE STATUS = 'R' AND TESTCASE_ID_FK = ""+tcIdPk+"""";
    stmt = connection.createStatement();
    rs = stmt.executeQuery(sql);
    while(rs.next()) {
    DefectTestCaseMappingDTO defectDTO= new DefectTestCaseMappingDTO();
defectDTO.setTestCaseId(getTcById(rs.getInt(""TESTCASE_ID_FK"")));
defectDTO.setStatus(rs.getString(""STATUS""));
defectDTO.setAuto(rs.getString(""AUTO""));	
defectDTO.setReviewed(rs.getString(""REVIEWED""));
fieldDefectsList.add(defectDTO);}}
			}
}
return con; 
\end{lstlisting}} & \RaggedRight{\begin{lstlisting}
public DataSet getTestCaseByTestCaseId(int testCaseId){
System.out.println('getTestCaseByTestCaseId');
Connection con = null;
Statement stmt = null;
ResultSet rs = null;
DataSet ds = new DataSet();
try{
    con = ConnectionProvider.getConnection();
    if(null != con){
        stmt = con.createStatment();
        String sql = "SELECT * FROM TESTCASES where TESTCASE_ID_FK="+testCaseId+"";

        while(rs.next()) {
            TestCaseDAO tcDao = new TestCaseDAO();
            tcDao.setTestCaseName(rs.getString("TESTCASE_NAME"));
            tcDao.setTestCaseDesc(rs.getString("TESTCASE_DESC"));
            tcDao.setTestCaseId(rs.getInt("TESTCASE_ID_FK"));
            ds.add(tcDao);
        }
        return ds;
        } catch (Exception e) {
    }}}
\end{lstlisting}}\\ \hline 
\end{longtable}
\end{center}

\pagebreak

\section{Guidelines and Recommendations}
The following guidelines and recommendations have been summarized below, based on the various experiements that we have done on text and code fine tuning:
\begin{itemize}
\item Empirically loading the model in half precision is sufficient to go ahead with fine tuning and it also saves GPU memory to accommodate more batches if needed to save on finetuning time
    \item Unless there is an abundance of data, parameter efficient finetuning is preferable than full finetuning. This also helps in creating easily moveable low sized adapters tuned for different tasks or domains
    \item Choose a model quantization level based on section \ref{cest}. For example, consider Llama 7B model in a 16 GB colab environment; in this scenario, 8 bit quantized LORA fine tuning is possible but not full model fine tuning.
  \item For full fine tuning, typically multiple GPUs are required. In case of a constraint of having only one GPU available and a large CPU memory, it is recommended to use paged adam optimizer 
  \item For small datasets, it is ideal to use LORA fine tuning. Rank and Alpha has to be fine tuned
   \item From the empirical experiments on text and code data, to make a language model assimilate new information, lower rank and higher alpha is recommended
    \item For large documents with text content of the order of few hundred MBs, it is recommended to utilise the full sequence length capability of the model in every row of data 
    \item Fine tuning time largely depends on the number of rows in dataset. If the text content is chunked to full context length without padding, number of data rows can be greatly reduced
  \item Gradient accumulation steps is the number of steps after which the optimizer is stepped. Until then gradients are accumulated over the batches. This is good in distributed system but in single GPU it is slow
  \item A higher batch size will lead to faster convergence and might give better performance at inference. Batch size is recommended to be kept at a lower value suitable for the model and not to the limiting value of GPU memory. 
  \item Higher the gradient accumulation steps more the memory will be saved but at the cost of longer fine tuning time.

\end{itemize}
\section{Further work}
 In this paper we show that LLMs are able to learn new information from limited data with right LORA configurations. However the results have traces of hallucinations. To mitigate hallucinations, within the current setting different prompt templates have to be experimented. It also boils down to the way dataset is prepared. Chunking techniques like semantic chunking provide a way to create chunks that stand on their own as separate information entities. This could be explored further as a dataset preparation recipe to reduce hallucinations.

\section{Conclusion}
The paper discussed on the topic of fine tuning open source Large Language Models with proprietary documents and code repositories. In the Dataset preparation sections detailed steps on creating the dataset from raw documents and code bases is given. It is followed by experiments with different methods of preparation and manual evaluation of the model responses with different LORA configurations. Finally some pointers observed during the fine tuning process are given as guidelines.

%Bibliography
\bibliographystyle{unsrt}  
\bibliography{references}

\begin{thebibliography}{10}

\bibitem{xu2024paradigm}
Haoran Xu, Young~Jin Kim, Amr Sharaf, and Hany~Hassan Awadalla.
\newblock A paradigm shift in machine translation: Boosting translation performance of large language models, 2024.

\bibitem{wang2024weaver}
Tiannan Wang, Jiamin Chen, Qingrui Jia, Shuai Wang, Ruoyu Fang, Huilin Wang, Zhaowei Gao, Chunzhao Xie, Chuou Xu, Jihong Dai, Yibin Liu, Jialong Wu, Shengwei Ding, Long Li, Zhiwei Huang, Xinle Deng, Teng Yu, Gangan Ma, Han Xiao, Zixin Chen, Danjun Xiang, Yunxia Wang, Yuanyuan Zhu, Yi~Xiao, Jing Wang, Yiru Wang, Siran Ding, Jiayang Huang, Jiayi Xu, Yilihamu Tayier, Zhenyu Hu, Yuan Gao, Chengfeng Zheng, Yueshu Ye, Yihang Li, Lei Wan, Xinyue Jiang, Yujie Wang, Siyu Cheng, Zhule Song, Xiangru Tang, Xiaohua Xu, Ningyu Zhang, Huajun Chen, Yuchen~Eleanor Jiang, and Wangchunshu Zhou.
\newblock Weaver: Foundation models for creative writing, 2024.

\bibitem{zheng2023building}
Zhonghua Zheng, Lizi Liao, Yang Deng, and Liqiang Nie.
\newblock Building emotional support chatbots in the era of llms, 2023.

\bibitem{touvron2023llama}
Hugo Touvron, Louis Martin, Kevin Stone, Peter Albert, Amjad Almahairi, Yasmine Babaei, Nikolay Bashlykov, Soumya Batra, Prajjwal Bhargava, Shruti Bhosale, Dan Bikel, Lukas Blecher, Cristian~Canton Ferrer, Moya Chen, Guillem Cucurull, David Esiobu, Jude Fernandes, Jeremy Fu, Wenyin Fu, Brian Fuller, Cynthia Gao, Vedanuj Goswami, Naman Goyal, Anthony Hartshorn, Saghar Hosseini, Rui Hou, Hakan Inan, Marcin Kardas, Viktor Kerkez, Madian Khabsa, Isabel Kloumann, Artem Korenev, Punit~Singh Koura, Marie-Anne Lachaux, Thibaut Lavril, Jenya Lee, Diana Liskovich, Yinghai Lu, Yuning Mao, Xavier Martinet, Todor Mihaylov, Pushkar Mishra, Igor Molybog, Yixin Nie, Andrew Poulton, Jeremy Reizenstein, Rashi Rungta, Kalyan Saladi, Alan Schelten, Ruan Silva, Eric~Michael Smith, Ranjan Subramanian, Xiaoqing~Ellen Tan, Binh Tang, Ross Taylor, Adina Williams, Jian~Xiang Kuan, Puxin Xu, Zheng Yan, Iliyan Zarov, Yuchen Zhang, Angela Fan, Melanie Kambadur, Sharan Narang, Aurelien Rodriguez, Robert Stojnic, Sergey Edunov, and Thomas
  Scialom.
\newblock Llama 2: Open foundation and fine-tuned chat models, 2023.

\bibitem{liu2023fingpt}
Xiao-Yang Liu, Guoxuan Wang, Hongyang Yang, and Daochen Zha.
\newblock Fingpt: Democratizing internet-scale data for financial large language models, 2023.

\bibitem{wu2023pmcllama}
Chaoyi Wu, Weixiong Lin, Xiaoman Zhang, Ya~Zhang, Yanfeng Wang, and Weidi Xie.
\newblock Pmc-llama: Towards building open-source language models for medicine, 2023.

\bibitem{app14052074}
Rajvardhan Patil and Venkat Gudivada.
\newblock A review of current trends, techniques, and challenges in large language models (llms).
\newblock {\em Applied Sciences}, 14(5), 2024.

\bibitem{Jeong_2023}
Cheonsu Jeong.
\newblock A study on the implementation of generative ai services using an enterprise data-based llm application architecture.
\newblock {\em Advances in Artificial Intelligence and Machine Learning}, 03(04):1588–1618, 2023.

\bibitem{cuconasu2024power}
Florin Cuconasu, Giovanni Trappolini, Federico Siciliano, Simone Filice, Cesare Campagnano, Yoelle Maarek, Nicola Tonellotto, and Fabrizio Silvestri.
\newblock The power of noise: Redefining retrieval for rag systems, 2024.

\bibitem{balaguer2024rag}
Angels Balaguer, Vinamra Benara, Renato~Luiz de~Freitas~Cunha, Roberto de~M.~Estevão~Filho, Todd Hendry, Daniel Holstein, Jennifer Marsman, Nick Mecklenburg, Sara Malvar, Leonardo~O. Nunes, Rafael Padilha, Morris Sharp, Bruno Silva, Swati Sharma, Vijay Aski, and Ranveer Chandra.
\newblock Rag vs fine-tuning: Pipelines, tradeoffs, and a case study on agriculture, 2024.

\bibitem{hu2021lora}
Edward~J. Hu, Yelong Shen, Phillip Wallis, Zeyuan Allen-Zhu, Yuanzhi Li, Shean Wang, Lu~Wang, and Weizhu Chen.
\newblock Lora: Low-rank adaptation of large language models, 2021.

\bibitem{dettmers2022llmint8}
Tim Dettmers, Mike Lewis, Younes Belkada, and Luke Zettlemoyer.
\newblock Llm.int8(): 8-bit matrix multiplication for transformers at scale, 2022.

\bibitem{NEURIPS2023_1feb8787}
Tim Dettmers, Artidoro Pagnoni, Ari Holtzman, and Luke Zettlemoyer.
\newblock Qlora: Efficient finetuning of quantized llms.
\newblock In A.~Oh, T.~Neumann, A.~Globerson, K.~Saenko, M.~Hardt, and S.~Levine, editors, {\em Advances in Neural Information Processing Systems}, volume~36, pages 10088--10115. Curran Associates, Inc., 2023.

\bibitem{6757323}
Mark Horowitz.
\newblock 1.1 computing's energy problem (and what we can do about it).
\newblock In {\em 2014 IEEE International Solid-State Circuits Conference Digest of Technical Papers (ISSCC)}, pages 10--14, Feb 2014.

\bibitem{kalamkar2019study}
Dhiraj Kalamkar, Dheevatsa Mudigere, Naveen Mellempudi, Dipankar Das, Kunal Banerjee, Sasikanth Avancha, Dharma~Teja Vooturi, Nataraj Jammalamadaka, Jianyu Huang, Hector Yuen, Jiyan Yang, Jongsoo Park, Alexander Heinecke, Evangelos Georganas, Sudarshan Srinivasan, Abhisek Kundu, Misha Smelyanskiy, Bharat Kaul, and Pradeep Dubey.
\newblock A study of bfloat16 for deep learning training, 2019.

\bibitem{jacob2017quantization}
Benoit Jacob, Skirmantas Kligys, Bo~Chen, Menglong Zhu, Matthew Tang, Andrew Howard, Hartwig Adam, and Dmitry Kalenichenko.
\newblock Quantization and training of neural networks for efficient integer-arithmetic-only inference, 2017.

\bibitem{infissue}
{Memory Decreases! But Latency Increases....}, howpublished = {\url{https://github.com/timdettmers/bitsandbytes/issues/6} }.

\bibitem{zhang2023instruction}
Shengyu Zhang, Linfeng Dong, Xiaoya Li, Sen Zhang, Xiaofei Sun, Shuhe Wang, Jiwei Li, Runyi Hu, Tianwei Zhang, Fei Wu, and Guoyin Wang.
\newblock Instruction tuning for large language models: A survey, 2023.

\bibitem{rake}
{Rapid Automatic Keyword Extraction algorithm} domain independent keyword extraction algorithm which tries to determine key phrases in a body of text by analyzing the frequency of word appearance and its co-occurance with other words in the text.
\newblock \url{https://pypi.org/project/rake-nltk/}.

\bibitem{llmsearch}
{Querying local documents, powered by LLM}.
\newblock \url{https://github.com/snexus/llm-search/blob/main/src/llmsearch/parsers/doc.py}.

\bibitem{gunasekar2023textbooks}
Suriya Gunasekar, Yi~Zhang, Jyoti Aneja, Caio César~Teodoro Mendes, Allie~Del Giorno, Sivakanth Gopi, Mojan Javaheripi, Piero Kauffmann, Gustavo de~Rosa, Olli Saarikivi, Adil Salim, Shital Shah, Harkirat~Singh Behl, Xin Wang, Sébastien Bubeck, Ronen Eldan, Adam~Tauman Kalai, Yin~Tat Lee, and Yuanzhi Li.
\newblock Textbooks are all you need, 2023.

\end{thebibliography}

\end{document}